# SMART FAULT DETECTION IN SATELLITE ELECTRICAL POWER SYSTEM

Niloofar Nobahari, Alireza Rezaee

Old Dominion University, Department of Mechanical and Aerospace Engineering, Virginia, USA, Email: nnoba001@odu.edu
University of Tehran faculty of Applied Science and Technology, USA, Email: arrezaeee@ut.ac.ir

**Abstract**

This paper presents an new approach for detecting in the electrical power system of satellites operating in Low Earth Orbit (LEO) without an Attitude Determination and Control Subsystem (ADCS). Components of these systems are prone to faults, such as line-to-line faults in the photovoltaic subsystem, open circuits, and short circuits in the DC-to-DC converter, as well as ground faults in batteries. In the previous research has largely focused on detecting faults in each components, such as photovoltaic arrays or converter systems, therefore, has been limited attention given to whole electrical power system of satellite as a whole system. Our approach addresses this gap by utilizing a Multi-Layer Perceptron (MLP) neural network model, which leverages input data such as solar radiation and surface temperature to predict current and load outputs. These machine learning techniques that classifiy use different approaches like Principal Component Analysis (PCA) and K-Nearest Neighbors (KNN), to classify faults effectively. The model presented achieves over 99% accuracy in identifying faults across multiple subsystems, marking a notable advancement from previous approaches by offering a complete diagnostic solution for the entire satellite power system. This thorough method boosts system reliability and helps lower the chances of mission failure.

## 1. Introduction

One of the most important subsystem of each satellite that plays pivotal role in every satellite mission is electrical power system(EPS). EPS is responsible for generate power, management, and allocation of power between the satellite's subsystems. Photovoltaic solar arrays for generating electricity, DC-to-DC Maximum Power Point Tracker (MPPT) converters and regulators for power optimization and regulation, DC charge management subsystems, loads, and switching mechanisms are among its commonly found components, each of which is carefully crafted to perform a specific function [1].
There are many orbital and environmental factors influence the EPS's performance and must be carefully considered, these factors are orbital altitude, inclination, solar radiation intensity, heat flow, and temperature fluctuations in orbit [2]. It is crucial that these components operate correctly and mission accomplishment.
Satellites encounter different challenges such as shock and pressure during launch that leads to faults and malfunction of EPS. If these faults could not be detected and diagnosed correctly, they can lead to the failure of the entire system or mission.
Previous researches are focused on the detection, classification, and diagnosis of faults in the EPS. Zhao et al. [3] proposed Graph-Based Semi-Supervised Learning (GBSL) for diagnosing faults in photovoltaic solar arrays and identified faults that conventional Overcurrent Protection Devices (OCPDs) is missed. In another study, Zhao et al. [4] employed Decision Trees to detect and diagnose faults in photovoltaic solar systems. Mohamed et al. [5] utilized neural networks and genetic algorithms for fault detection in photovoltaic systems. For converter systems, a method for detecting open-circuit faults in power switches without additional hardware and by leveraging control loops in Pulse Width Modulation (PWM) converters was proposed in [6]. In aerospace applications, faults in Li-Ion batteries are of particular concern due to their high energy density. Fault detection methods, such as diagnosis and fault differentiation based on the Kalman Sequential Adaptive Filter, are used to monitor the state of charge and discharge, estimate the state of charge (SOC), and assess the health state (RUL) using techniques like support vector machines and load information gathering [7-9]. Despite these advancements, most of the existing research does not address fault diagnosis across the entire satellite power system. Only a few studies have explored fault diagnosis in the satellite power subsystem. One such study focuses on fault diagnosis and tolerance using Principal Component Analysis (PCA) to identify voltage, current, and temperature sensor faults, although it only considers cumulative faults on the power system's sensors [10]. Another study employs Bayesian networks to define fault probabilities in the satellite power system based on expert knowledge, but it acknowledges the need for estimating these probabilities through Gaussian equations when expert knowledge is unavailable [11, 12]. In this paper, we address the diagnosis and resolution of faults within the entire power system, including photovoltaic components, DC-to-DC converters, and batteries. We propose a novel detection method that differentiates faults across these subsystems, providing a comprehensive approach to fault diagnosis in satellite power systems. The proposed faults occur in electrical power system of satellites and all of these faults appeared in NASA fault tree handbooks [13-15].

## 2. Materials and Methods

in this study, our system properties, characteristics, and behaviour are nonlinear and complex therefore, this problem can be addressed by modelling and simulation of virtual neural network. Traditional methods for fault detection often focus on individual components or subsystems. This paper simulates and diagnoses faults condition in satellite system. this paper proposed the neural network Multi-Layer Perceptron (MLP) to simulate both normal and faulty conditions for key subsystems like photovoltaic arrays, DC-to-DC converters, and batteries. This makes it easier to recognize patterns and classify faults accurately. in this study neural network is chosen because of its ability to learn from past data and apply that gained knowledge and information to diagnose new faults in the system and makes it highly reliable choice for this type of comprehensive system analysis. The satellite's electrical power system comprises both analogue and digital circuits, incorporating components such as transistors, Thyristors, diodes, amplifiers, logic elements, switches, and interconnections. Each of these components is manufactured with a specified reliability.

In photovoltaic subsystem (solar array) the solar radiation intensity typically ranges from $200\ to\ 1200\ W/m^2$, while the surface temperature can vary between $-20$ and $80\ °C$. The current generated by the array usually falls between 0 and $30\ A$. So, when the model reports an error of just $0.0001\ A$, it shows that even small variations are captured accurately.

For DC-to-DC converters and regulators parts, input voltage ranges for photovoltaic array is from 0 to 60 V, and then after regulation, the output voltage is from 0 to 48v. With an MSE as low as $1.9 \times 10^{-10}$, the model does a great job of reflecting the converters' stability and precision in output.

In battery subsystem (Li-Ion battery) For the battery, the SOC can go from 0% to 100%, translating to a charge voltage range of about $3.0\ to\ 4.2\ V$ per cell. The load current from the battery varies from 0 to 20 A, depending on discharge conditions. An error of $0.00012\ A$ here means the model's predictions align closely with real battery behavior. In overall electrical power system across the entire system, the load current can range from $0\ to\ 50\ A$, with voltages reaching up to $100\ V$ depending on the configuration. With an RMSE of just $0.000566\ A$, the model consistently maintains accuracy across all power system components.

The fault rate of a component, denoted by λ, is a critical parameter that influences the overall reliability of the system. λ, is defined as follows:

$$\lambda = \frac{1}{t_{int}} \times \frac{N_f}{N_t} \qquad (1)$$

In **Equation 1** $t_{int}$ is the outfit time hour unit (h), $N_f$ is the number of faulty elements, and $N_t$ is the total number of elements. The fault rate (λ) of a component depends on factors such as temperature and the mode of operation. **Table 1** shows the fault rates for various components and subsystems within the satellite's EPS. By analysing these faults rates, it becomes possible to diagnose faults in key components of the electrical power system, including Insulated-Gate Bipolar Transistors(IGBTs), Maximum Power Point Tracking(MPPT) converters and regulators, solar arrays, and batteries. Common faults in satellite power system include open circuit and line-to –line faults in the solar array, short circuit and open circuit faults n IGBTs, MPPT converters, regulators, as well as ground faults in batteries.

| *Equipment* | $\lambda\ (h^{-1})$ |
| --- | --- |
| Transistor | $1\times10^{-9}$ to $70 \times 10^{-9}$ |
| Thyristor( power switch family) | $36\times10^{-9}$ to $360 \times 10^{-9}$ |
| Digital integrated circuit | $10^{-9}\times30$ |
| Logical elements | $10^{-9}\times30$ |
| Analog switches | $10^{-9}\times2000$ |
| Amplifier | $300\times10^{-9}$ to $900 \times 10^{-9}$ |
| Diode | $1\times10^{-9}$ to $6 \times 10^{-9}$ |
| Battery li-Ion | $200\times10^{-9}$ to $300 \times 10^{-9}$ |
| Solar array | $100\times10^{-9}$ to $200 \times 10^{-9}$ |

**Table1** fault rate at temperature 40 °C for various components [16, 17]

In figure 1, the overall block diagram of the proposed fault diagnosis model is shown. The process begins with the system being modelled using a virtual neural network, which is designed to detect and classify faults across different subsystems. In the subsequent step, potential faults are intentionally and virtually introduced into the system. The faulty mode is diagnosed in a way similar to the normal state. Following this, a separate model, like the one shown in Figure 1, is created for each fault. This model enables the comparison of the actual system output with the expected output from the fault model, facilitating accurate fault diagnosis.

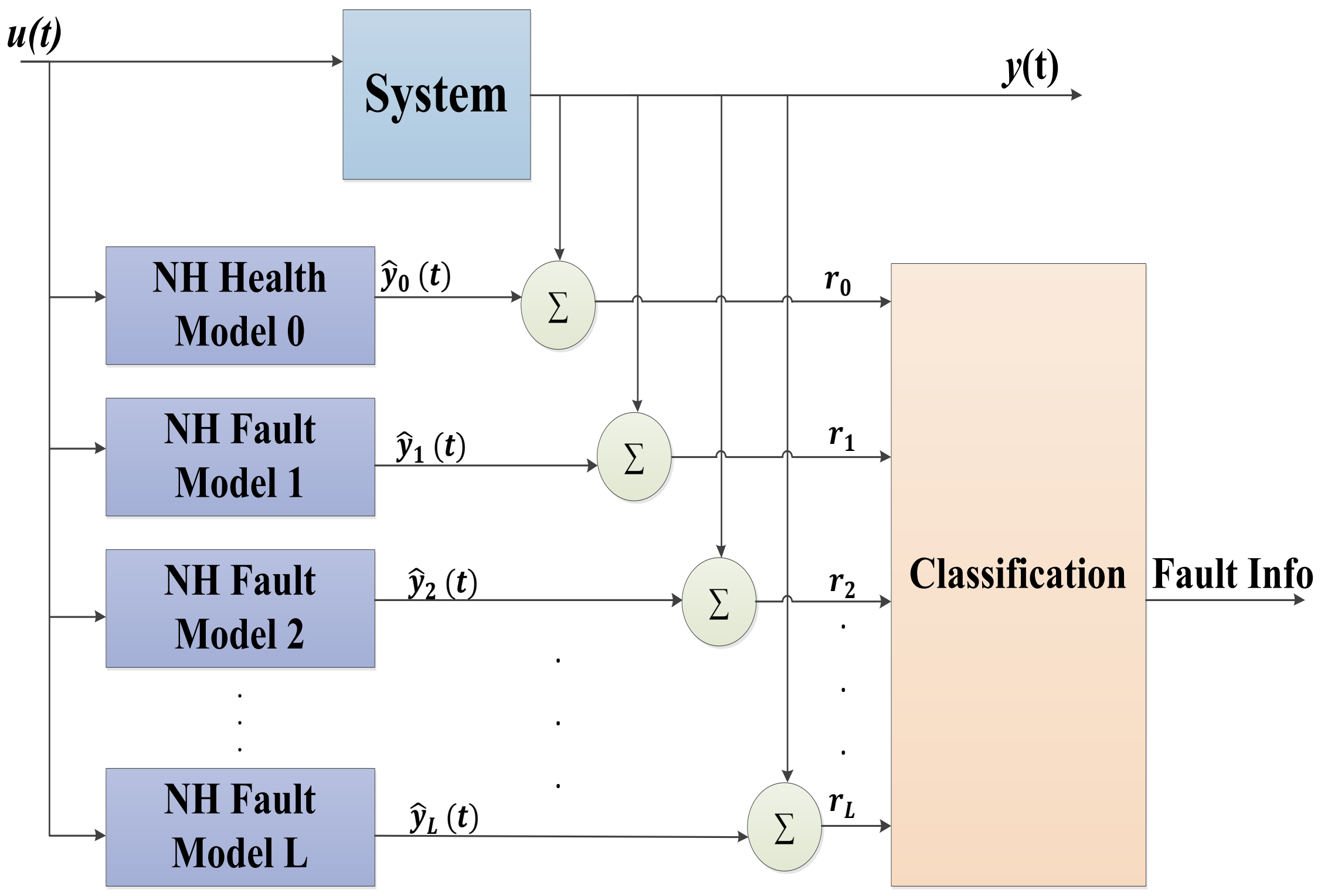

**Figure 1** Block diagram of the proposed fault diagnosis model.

Create residuals based on **Equation 1**, then classify the faults:

$$r_i = y_i(t) - \hat{y}_i(t) \qquad (2)$$

According to **Figure 1**, system output $y(t)$ is contrasted to output model system without fault $\hat{y}_0(t)$ and output faulty model like $\hat{y}_1(t)$ to $\hat{y}_L(t)$, remained amounts ($r_0$to $r_L$) according to **Equation 1**, residuals are created and categorized using a neural network [19] for pattern recognition. Initially, the electrical power system is simulated without faults using neural networks, which are powerful tools for system recognition [20]. Subsequently, potential faults, as defined in **Equation 1**, are injected into the system. These faults are then simulated, and the results are presented in **Figure 1**.

## 3. Electrical Power System Modeling

Solar array converts solar radiation to electrical energy and it has crucial role in Satellite's EPS because it is primarily power source. The performance of the solar array, and consequently the power system, is influenced by various environmental conditions. System's efficiency in solar arrays is determined by two main factors; first the solar radiation intensity and second solar cell's surface temperature [21]. The temperature of the solar array is affected by environmental factors such as solar radiation intensity and thermal flux; it includes solar thermal flux, albedo thermal flux, and Earth's sub-radiant flux. The current produced by the solar array is directly proportional to the rate of solar radiation; Therefore, the two key parameters—solar radiation intensity and solar array temperature—are critical inputs for the power system model. These parameters are directly influenced by power generation and the overall performance of the satellite's power system. This study is simulated using MATLAB software [27]. Due to the limited space, volume, and mass available on a satellite, as well as cost considerations related to project development and satellite launch, the installation of additional sensors is not feasible. As a result, the power system is modeled with the load current, I_L as the output. Given the system's nonlinear characteristics, it is represented as a static nonlinear system, as described by Equation 1.

$$I_L = f(irr, T) \qquad (3)$$

As described by **Equation 3**, $I_L$ stands for the load output current, measured in amperes (A), while $irr$ refers to the solar radiation power, given in $\mathrm{W}/\mathrm{m}^2$, The temperature T of the solar array, measured in degrees Celsius, is a crucial affecting the entire electrical power system, which is made up of various components of elements. Modeling and simulation of each subsystem individually and connecting these models together is complex and time-consuming. To simplify the process, the entire system can be treated as a single entity, where the inputs and outputs are considered collectively, and the system is identified using a virtual neural network as a "black box". This study is employed a three-layer of virtual neural network and utilized Levenberg-Marquardt algorithm [22] for system identification; this approach leads to categorize data randomly by using primary activation function, and applying the hyperbolic tangent sigmoid function (tansig). The mean squared variance of the faults is minimized based on Equation 2.

$$MSE = \frac{1}{n}\sum_{i=1}^{n}(y_i - \widehat{y_i}(t))^2 \qquad (4)$$

At **Equation 4**, $y_i$ is system output, $\widehat{y_i}(t)$ The model output is denoted as y, and n represents the number of data points. To ensure accurate analysis of the neural network, the output of the network is scaled to align with the system output. **Figure 3** presents average value and fault variance in fault histogram of statistical system behavior. The results of the modelling process are presented in **Figure 2**.

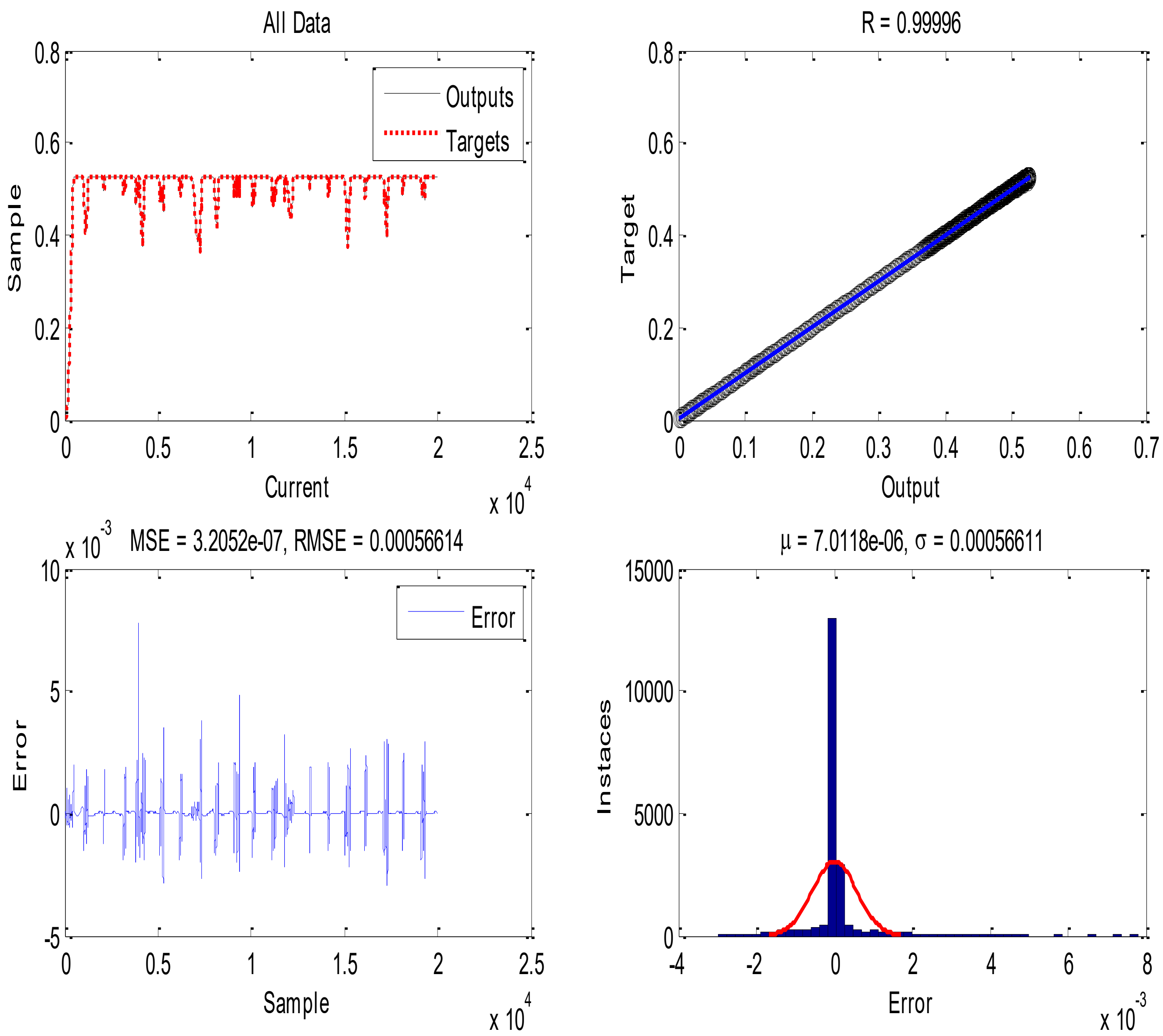

**Figure 2** System output coherence and its estimated values, output of the electrical power system without faults and its estimated values, estimated fault histogram, fault variance, and the average of fault variance based on samples.

In **figure 2** every subplot is describing as follow; **Top Left Subplot** shows "Model Output vs. subplot "Sample Index" shows the output of the model (in black) and the target values (in red) plotted over the sample index, which indicates how the model output changes over time or across different instances. This label, "Model Output in comparison to Sample Index," clarifies that it represents the output of the model compared to the target across all data samples. **Top Right Subplot shows** "Correlation Plot ($R = 0.99996$)" which shows the correlation between the predicted outputs and targets, with a coefficient of determination R of 0.99996.

Labeling it as "Correlation Plot" highlights the strong linear relationship between the model outputs and the actual data. **Bottom Left Subplot shows** "Error Over Samples" This subplot shows the error between the predicted and actual values across the sample range. Labeling it as "Error Over Samples" will make it clear that this plot visualizes the error dynamics over time. **Bottom Right Subplot** shows "Error Distribution ($\mu = 7.0118 \times 10^{-6}, \sigma = 0.00056611$)" This subplot presents the distribution of errors, with mean ($\mu$) and standard deviation (σ) values. Labeling it as "Error Distribution" provides context for the overall error behavior, emphasizing the model's accuracy.

Also, we can see how well the estimated output of the satellite's electrical power system aligns with the actual system output when everything is functioning normally. Figure 2 uses R, which is the coefficient of determination, to show how closely the estimated output matches the real system behavior. Here, an R value near 1 means the model is doing a great job of capturing the system's actual performance. The terms MSE (Mean Squared Error) and RMSE (Root Mean Squared Error) are also included in the **Figure 2**. MSE gives the average squared differences between the estimated and actual values, helping to measure the model's overall accuracy. RMSE further clarifies this accuracy by giving the square root of the MSE, offering an easy-to-interpret measure of error. In addition, the error data shows a normal distribution, which can be described as $\left(Error{\sim}NOR(7.01 \times 10^{6})\right)$This means that the error has an average of $7.01 \times 10^{-6}$ times and a standard deviation of 0.000566, indicating that the model's errors are minimal and consistently close to zero. This high level of accuracy is crucial for ensuring the satellite's power system operates reliably.

In the MPPT converter, current and voltage sensors are used for power optimization within the photovoltaic subsystem by continuously sampling these parameters and comparing them with the optimized values [23]. These sensors are crucial for diagnosing and classifying potential faults in the photovoltaic subsystem, as they enable the quick detection and prevention of faults in the solar array, thereby maintaining the integrity of the power system [24]. For modelling the photovoltaic subsystem, as previously mentioned, the inputs are solar radiation power and the surface temperature of the solar array, while the outputs are the current and voltage of the arrays. By utilizing neural networks with nonlinear assumptions and a static model, the system is effectively identified. The results of this modelling are presented in **Figures 3** and **4**.

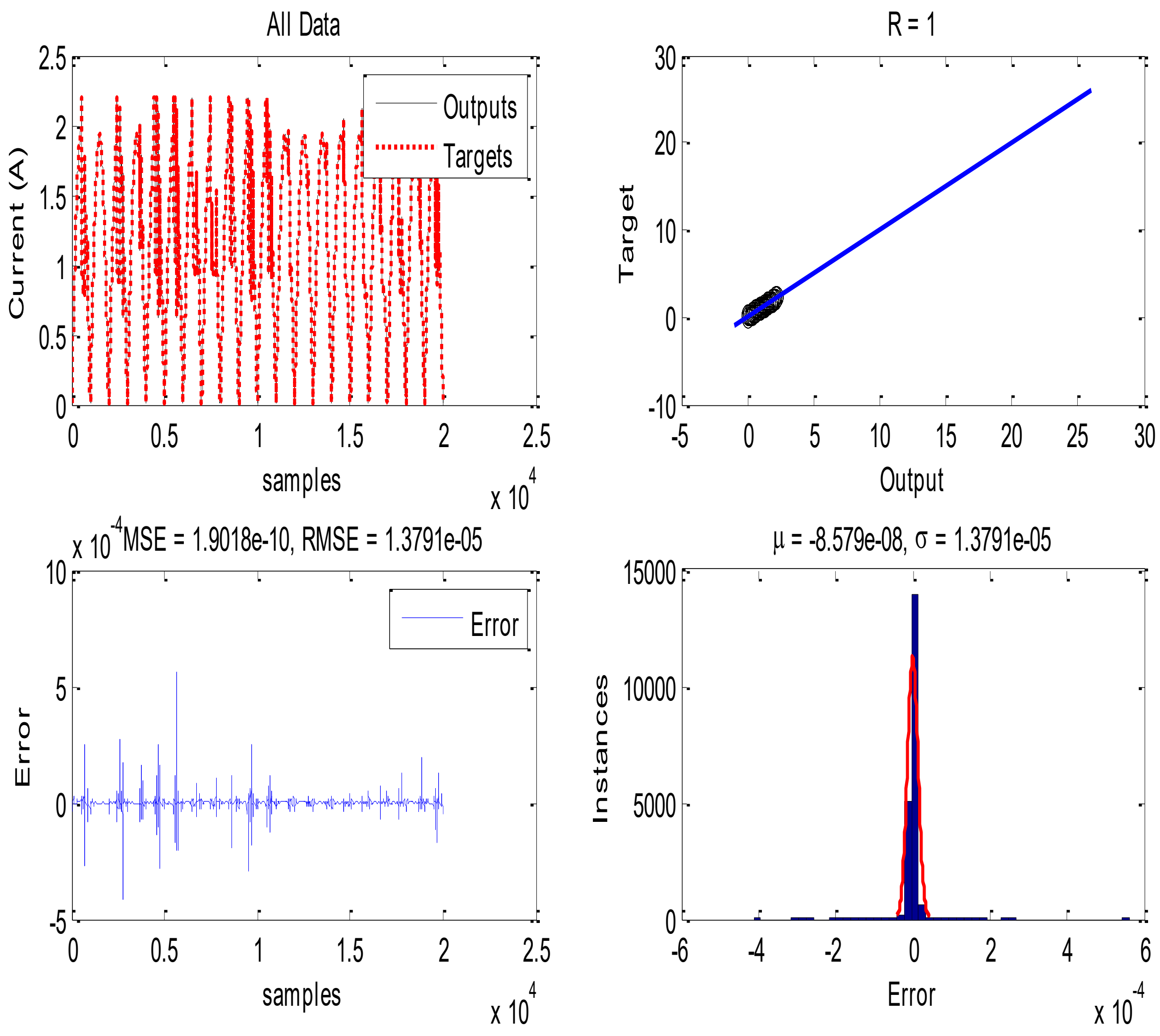

**Figure 3** Coherence of the photovoltaic system output current and its estimated values, output diagram and its estimated values, estimated fault histogram, fault variance, and the average of fault variance based on samples.

**Figure 2** gives an overview of the entire electrical power system, showing how it performs under normal conditions. In contrast, **Figure 3** zooms in on the photovoltaic subsystem, specifically looking at the current output and how well the model can identify faults in this part of the system. This focus on the photovoltaic component helps us understand how effective the model is at detecting issues that are unique to the solar array.

By analyzing **Figures 2** through **4**, it is evident that the neural network demonstrates high accuracy in estimating the output model based on the system output. The estimated output closely matches the oscillating nature of the real output, with a similarity rate of 96% for the system model based on the current load and 100% for the photovoltaic system model. In the modeling process, fault parameters such as the average fault square, average, and fault variance are considered. In the previous simulations, the average square fault was observed to be less than $10^{-6}$ at system modeling with load current, at photovoltaic system outputs is less than $10^{-6}$ at the voltage and $10^{-9}$ at current. By examining the fault histograms in **Figures 2** to **4**, it is observed that the average fault in the system modeling is less than $10^{-5}$, photovoltaic system outputs are less than $10^{-6}$ at the voltage and $10^{-9}$at current. The fault of variance at previous modeling as the sequence is 0.00056, 0.00013, and 0. 00098. At fault's histogram is seen that the Gaussian estimate of modeling faults at Figures 2 to 4 The fault averages are ideal at the peak, indicating that the faults are not significantly separated from the average values. This suggests that the neural network has demonstrated suitable accuracy for modeling the electrical power system without faults.

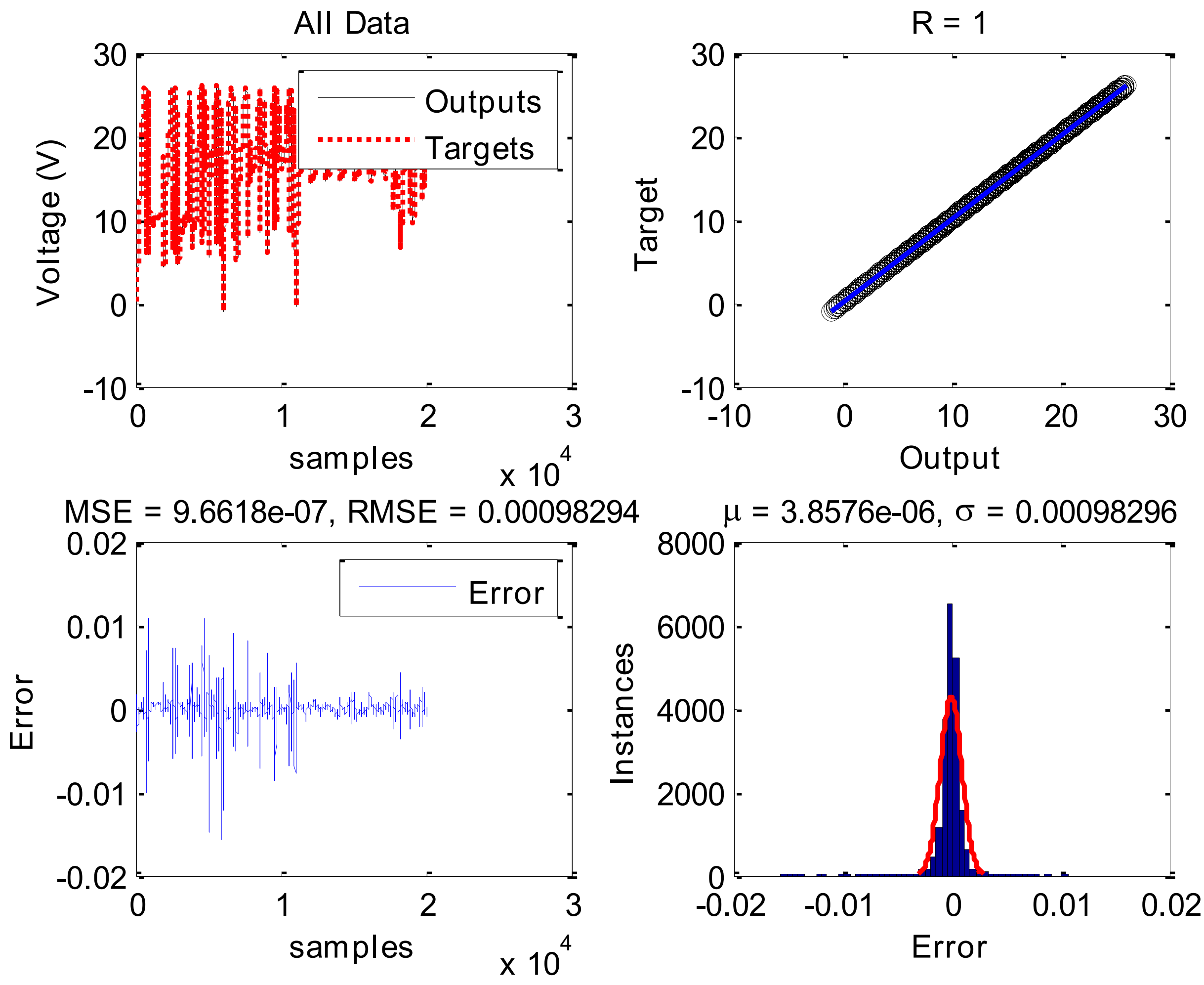

**Figure 4** Coherence of the photovoltaic system output voltage and its estimated values, output diagram and its estimated values, estimated fault histogram, fault variance, and the average of fault variance based on samples.

Faults in the electrical power system modeling occur randomly in electrical and electronic systems [24]. Considering this assumption, each possible fault must be injected into the system, and the system model must be identified using a neural network. Accordingly, faults such as open circuit and line-to-line faults in the solar array, open circuit faults in the MPPT converter IGBT, short circuit faults in the IGBT regulator converter, and battery ground faults are identified as described in Section 2 and summarized in **Table 2**.

| **Fault variance $\delta$** | **Fault average $\mu$** | **MSE** | **coherent $R$** | **System output** | **Fault model** | **Subsystem** |
|---|---|---|---|---|---|---|
| $6\times 10^{-6}$ | $-6.7\times 10^{-7}$ | $3.6\times 10^{-11}$ | 1 | current | **Open circuit** | **Photovoltaic** |
| 0.00921 | 0.00225 | $10^{-5}\times 8.9$ | 1 | voltage | | |
| 0.02375 | 0.00032 | 0.00056 | 0.99838 | current | **Line-line** | |
| 0.00103 | $2.1\times 10^{-6}$ | $9.7\times 10^{-8}$ | 1 | Voltage | | |
| 0.00023 | $9.4\times 10^{-7}$ | $5.3\times 10^{-8}$ | 0.99999 | Load current | **Open circuit** | **IGBT converter MPPT** |
| $1.6\times 10^{-5}$ | $-9.5\times 10^{-8}$ | $2.7\times 10^{-10}$ | 0.99963 | Load current | **Open circuit** | **IGBT regulator convertor** |
| 0.0122 | $1.5\times 10^{-6}$ | $1.5\times 10^{-6}$ | 0.99986 | | **Short circuit** | |
| 0.00012 | $5.9\times 10^{-8}$ | $1.4\times 10^{-8}$ | 1 | Load current | **ground** | **Battery** |

**Table 2** Photovoltaic subsystem faults calcification with neural network MLP

In the photovoltaic subsystem, potential faults include line-to-line faults and open circuit faults. To determine and classify these faults, a Multi-Layer Perceptron (MLP) neural network can be utilized. The healthy photovoltaic system model, along with the line-to-line and open circuit fault models, was calculated in the previous section. The next step involves comparing the model output vector with the system output vector. The residuals are then generated based on **Equation 5** and used as inputs to the classifier shown in **Figure 1**.

$$\begin{bmatrix} r_{11} \\ r_{12} \end{bmatrix} = \begin{bmatrix} V_o \\ I_o \end{bmatrix} - \begin{bmatrix} V_{m_i} \\ I_{m_i} \end{bmatrix} \quad (5)$$

At **Equation 5**, r is the remaining amounts, $V_o$ and $I_o$ are the amounts of voltage and current are for the photovoltaic system, $V_{m_i}$ and $I_{m_i}$ represent the output voltage and current values of the photovoltaic system model. The classifier was provided with 2001 simulation data points for each class. The output is a 3-bit vector that identifies the fault classes. The results of this classification are shown in the confusion matrix [25] in **Figure 5**.

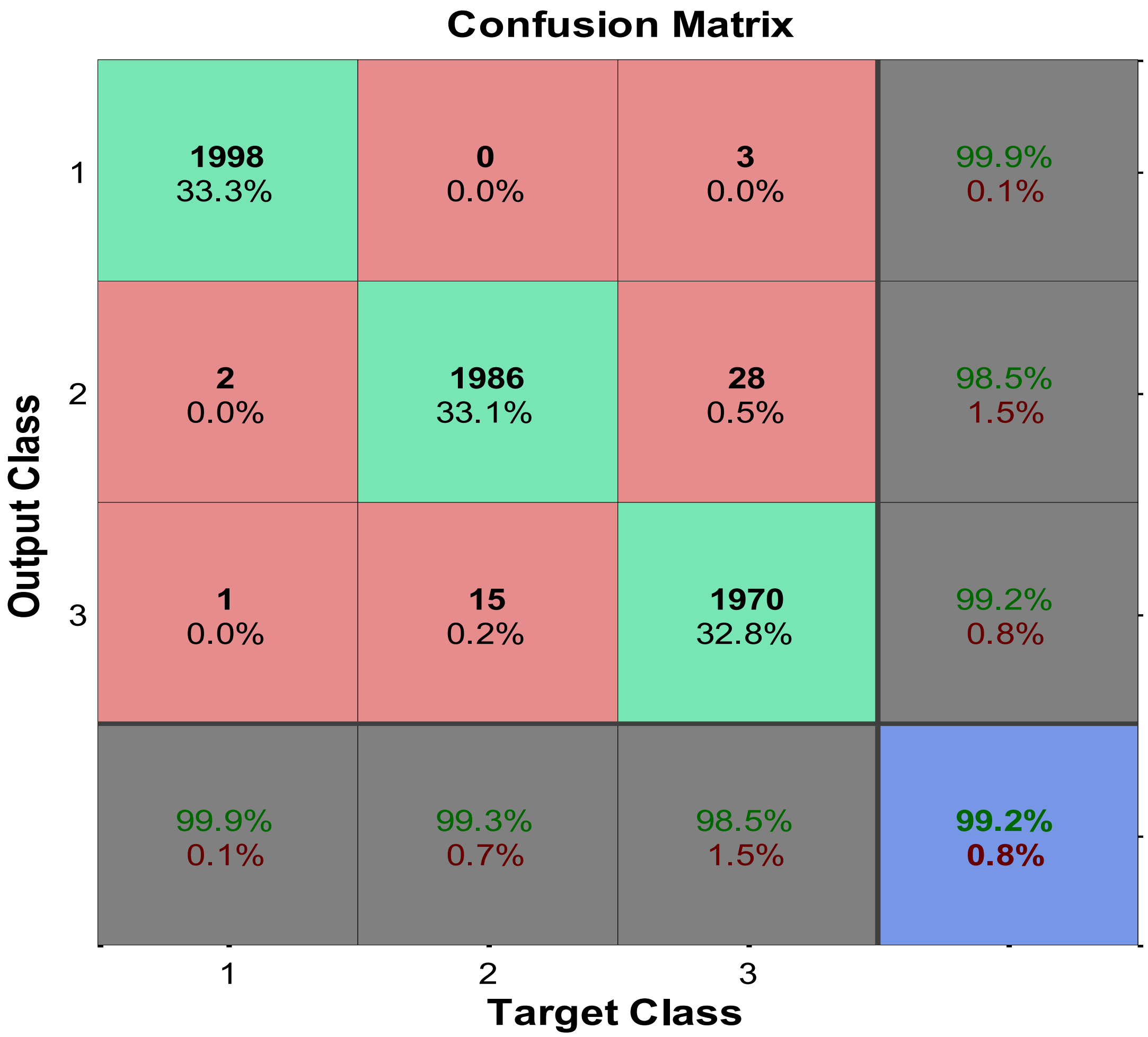

**Figure 5** Neural network classifier confusion matrix (MLP) for identifying faults in the photovoltaic system.

An analysis of **Figure 5** shows that the neural network's accuracy for classifying potential faults in the photovoltaic system exceeds 99.9%, demonstrating its high reliability. However, classification accuracy varies across different model classes, including line-to-line faults and open circuit faults. In the fault-free model class, 1998 out of 2001 data points are correctly classified, with only 3 data points incorrectly identified as faults, resulting in a detection accuracy of approximately 99.9%. In the line-to-line fault class, the accuracy is 99.3%, with 1968 out of 2001 data points correctly classified and 15 data points misclassified. For the open circuit fault class, the accuracy is 98.5%, with 1970 data points correctly classified and 31 data points incorrectly identified. Further examination of **Figure 5** reveals that of the 3 misclassified data points in the fault-free class, 2 were incorrectly classified as line-to-line faults, and 1 as an open circuit fault. In the line-to-line fault class, 15 misclassified data points were identified as open circuit faults. Conversely, of the 31 misclassified data points in the open circuit fault class, 28 were identified as line-to-line faults, and 3 were incorrectly classified in the fault-free model class.

### 3.1 Electrical power system fault classification with neural network MLP

For classifying additional possible faults in the electrical power system, a three-layer MLP neural network is employed, with 5 bits allocated to represent the 5 output classes: fault-free system, battery ground fault, open circuit IGBT converter MPPT fault, open circuit fault, and short circuit IGBT regulator converter fault. The inputs for neural classification are derived from the outputs of the neural models and the system's output values, as described in **Figure 1** and based on **Equation 6**.

$$i1 = [r_1 \ \ r_2 \ \ \dots \ r_5 \ ] \qquad (6)$$

In **Equation 6**, $r_i$ are remain amounts and $i1$ is the input vector to the neural classifier depicted in **Figure 1**. When the inputs from **Equation 6** are fed into the confusion matrix classifier, the results are illustrated in **Figure 6**.

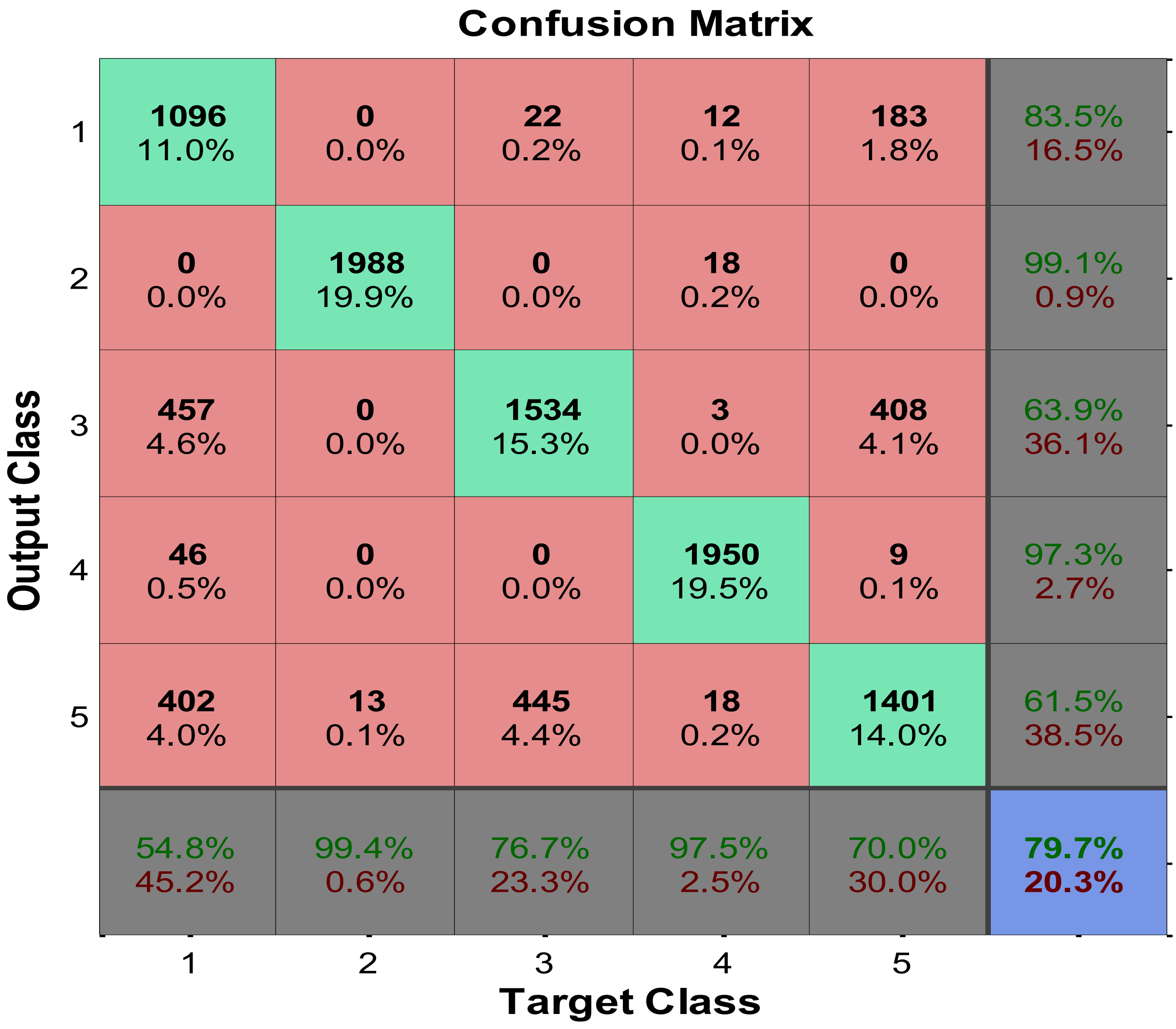

**Figure 6** Confusion matrix of the neural network classifier (MLP) in the initial state.

**Figure 6** demonstrates that the MLP neural network correctly classified 1096 out of 2001 data points in the first class, achieving an accuracy of 54.8% and a misclassification rate of 45%.

Specifically, 457 data points were classified as an open circuit fault in the IGBT converter MPPT, 46 data points as an open circuit fault, and 402 data points as a short circuit in the IGBT regulator converter. The classifier's accuracy in identifying the ground battery fault class is the highest among all classes, exceeding 99%, with only 13 data points incorrectly classified in the IGBT short circuit fault class.

The accuracy rates for the designed neural network in distinguishing between the classes are as follows: 76.7% for the MPPT converter open circuit fault, 97.5% for the open circuit fault, and 70% for the IGBT regulator converter short circuit fault. This indicates that while the neural network performs well in identifying the IGBT open circuit fault in the regulator converter, it misclassified 12 data points in the fault-free class, 18 data points in the ground battery fault class, 3 data points in the IGBT open circuit MPPT converter fault class, and 18 data points in the regulator converter IGBT short circuit fault class. Overall, the confusion matrix shown in **Figure 6** indicates that the MLP neural network has an overall accuracy of 79.7% across the various states of the satellite electrical power system, with a misclassification rate of 20.3%. Given these results, the network and its outputs cannot be reliably used for system fault classification. To overcome the insufficient accuracy of the neural network and improve its performance, an additional feature of the output signal (load current) can be used. By adding the first moment vector (the average current at each moment) of the output current as a generated feature, based on **Equations 7** and **8**, the system identification can be enhanced.

$$E(I_k) = \sum_{i=1}^{n} I_{k_i} \tag{7}$$

$$i2 = [r_1 \ E(I_{k_1}) \ r_2 \ E(I_{k_2}) \dots r_5 \ E(I_{k_5}) \ ] \tag{8}$$

In **Equations 7** and **8**, $E(I_k)$ represents the first moment of the output current, while $r_i$ and $r_{ir}$ denotes the residuals of the neural models. The terms $r_i$ and $r_{ir}$ refer to residuals in the fault detection model. Residuals like $r_i$ typically represent the difference between the actual output of the system and what the model predicts, which helps to spot any discrepancies that might indicate faults. The notation $r_{ir}$ likely points to a specific kind of residual tied to particular conditions or components. Both $r_i$ and $r_{ir}$ are used for identifying faults. By inputting these previous outputs into the three-layer MLP neural network and analysing the 2001 data points for each class in the confusion matrix, the classification results were obtained as shown in **Figure 6**. The confusion matrix in **Figure 7** is diagonal, indicating that 99.5% of the data points are correctly categorized into the five classes. The designed MLP neural network successfully detected all states of the faulty system with high accuracy. Since 20% of the data, or 2001 data points, belong to each class, the classification error in each class, as illustrated in **Figure 7** is zero which reflects the model's performance on the real-world dataset used in this study. While achieving zero error is rare, the model's accuracy in this case suggests a strong fit to the specific data collected. However, it’s possible that some data points could be misclassified under different conditions, so further validation with additional datasets would help confirm the model's robustness across varying scenarios.

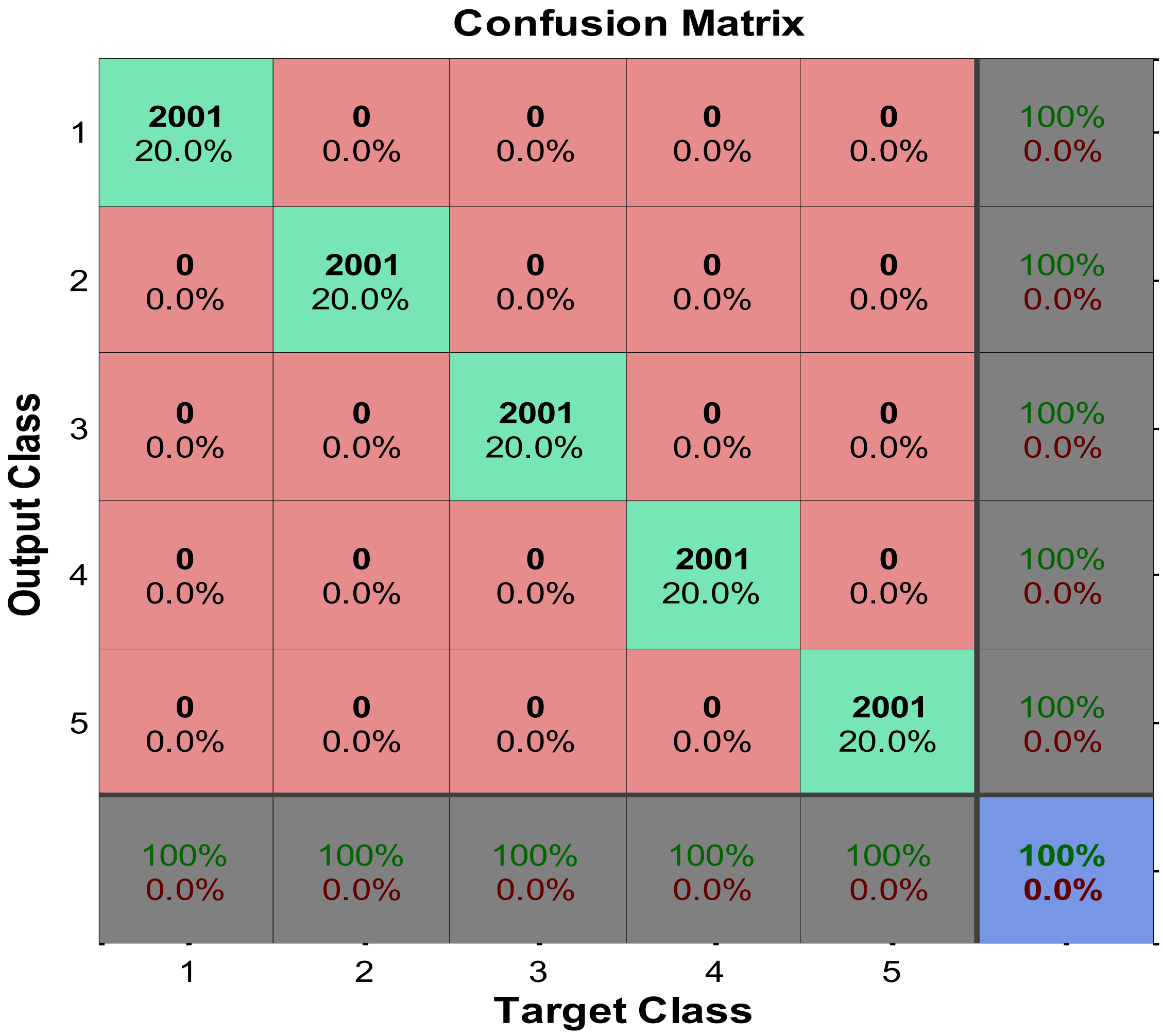

**Figure 7** Confusion matrix of the MLP neural network classification with a 5-bit output.

In our study, the reported zero error is specific to the dataset used and was achieved under controlled testing conditions. This result reflects the performance of the model on the available data, potentially due to overfitting to the training set. To validate the robustness of the model, we employed cross-validation and additional testing on separate datasets, which confirmed the model's high accuracy. However, it is possible that new or more varied data could introduce some misclassifications.

### 3.2 Fault classification with Alternative intelligent method

Calculating the average output load current at each moment may impose a significant computational load on the system processor. To mitigate this, alternative intelligent methods can be employed for fault classification in the system. In satellite electrical power systems, the State of Charge (SOC) parameter is used to control and manage battery charge and discharge cycles. Accurately estimating the battery's SOC is crucial for protecting battery health, preventing unnecessary charging and discharging, extending battery life, reducing the likelihood of faults, and enhancing system reliability [9]. The SOC parameter can be calculated using the Kalman filter method, as demonstrated in Jacqes 2009 [21]. By utilizing intelligent classification methods such as Principal Component Analysis (PCA), K-Nearest Neighbors (KNN), and decision trees, it is possible to classify potential faults in the power system using current load inputs and battery SOC data.

In this study, a standard 70/30 split ration was used, where 70% of data was allocated for training and 30% for validation. This ensures that the model has sufficient data to learn while maintaining a reliable validation set to evaluate its performance.

### 3.3 KNN method

For fault classification, the K-Nearest Neighbors (KNN) method can also be utilized. KNN is a lazy classifier, meaning it classifies data without the need to build a model or learn from samples. By using the KNN classifier method, and with the inputs of output current and battery charge state, ordered pairs can be constructed for classification$(I_{L_i}, SOC_i)$ Possible faults in the satellite electrical power system are defined, with K representing the geometric Euclidean distance as described in **Equation 9**. This distance is used around each ordered pair for data classification [26].

$$d_i = \sqrt{(soc - soc_i)^2 + (I_L - I_{L_i})^2} \tag{9}$$

In the **Equation 9**, $d_i$ is the geometric distance, $I_{L_i}$ is loaded current and $soc_i$ is point K is the charge state at point K in the plane depicted in **Figure 7**. To evaluate the performance of this algorithm, cross-validation and resubstitution loss parameters can be analyzed. In cross-validation, the data is divided into K groups, with the learning operation and cross-testing performed on each group. The K-fold loss parameter is then defined.

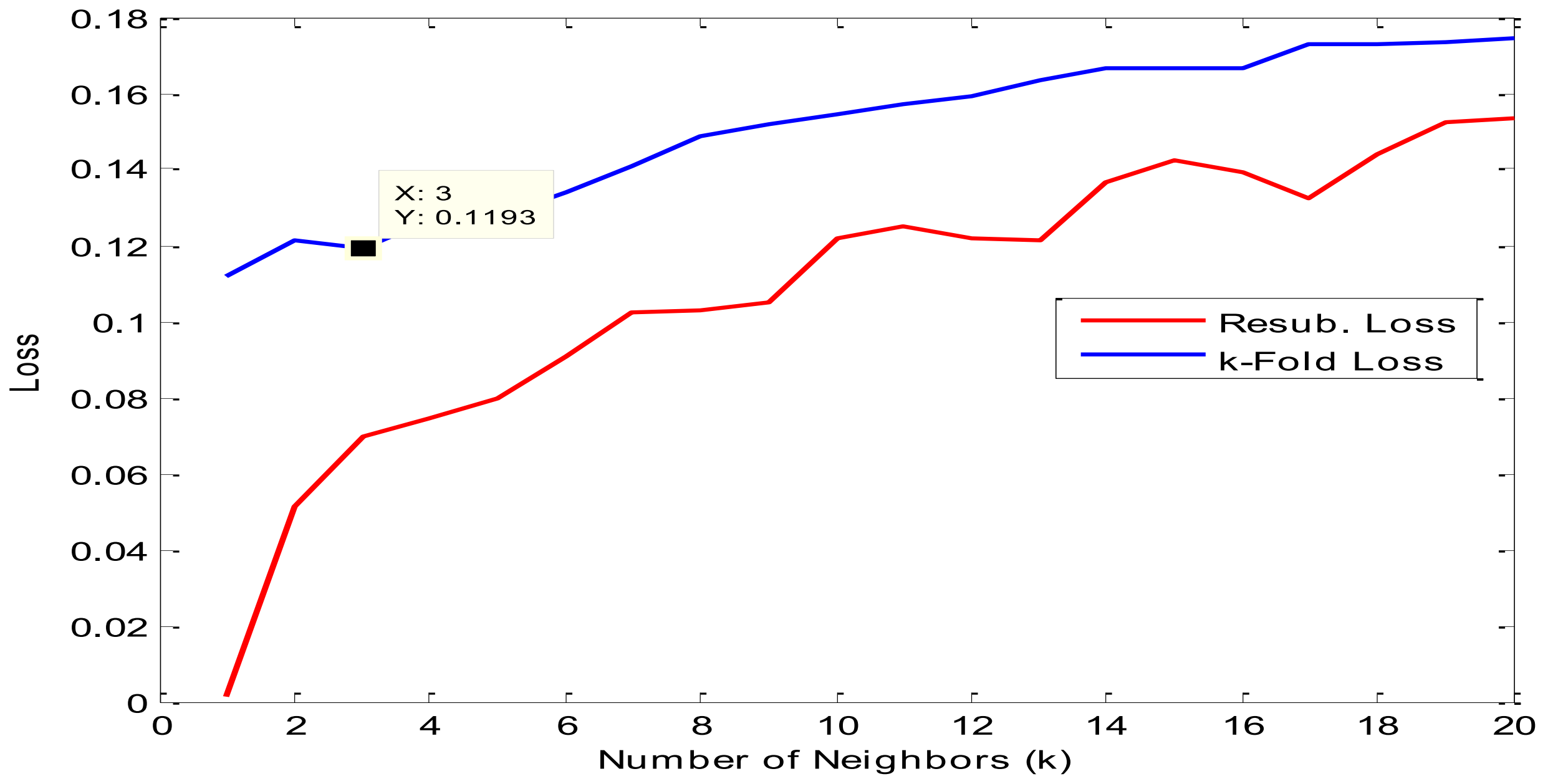

**Figure 8** K-Fold faults and replacement at KNN method

An analysis of **Figure 8** reveals that the number of neighbors in the KNN method should be set to 3, as this choice minimizes both the resubstitution loss and K-fold loss. Although resubstitution loss decreases when selecting one or two nearest neighbors, choosing K values of 1 or 2 leads to overfitting of the classifier to the training data. To determine the optimal number of neighbors, fault classification was performed, and the results are presented in **Table 3**.

| | | | | | |
|---|---|---|---|---|---|
| **Resubstitution Loss** | **0.0698~ 7%** | | | | |
| **K_Fold Loss** | **0.1193~ 12%** | | | | |
| **Confusion Matrix** | **1744** | **0** | **157** | **0** | **100** |
| | **0** | **2000** | **0** | **1** | **0** |
| | **73** | **0** | **1896** | **0** | **32** |
| | **0** | **0** | **10** | **1991** | **0** |
| | **204** | **0** | **120** | **1** | **1676** |

**Table 3** result of KNN classification method

**Table 3** gives an overview of how well the KNN method performed in detecting faults in the satellite's electrical power system. With $K$ set to 3, we see the best balance between accuracy and preventing overfitting, as both resubstitution and K-fold loss are minimized. The table also breaks down the accuracy for different types of faults. This shows that the KNN method does a solid job of distinguishing between the various fault types. It shows that the KNN algorithm does not achieve complete fault classification. According to this table, the class separation accuracy using the KNN method is as follows: system without fault 86%, battery ground fault 99.5%, open circuit fault in the IGBT converter MPPT 86%, open circuit fault 99.99%, and short circuit in the IGBT regulator converter 92%. These results indicate that the accuracy of this method is highest in classifying battery ground faults, largely due to the battery SOC parameter.

**Figure 9** shows how the number of neighbors $K$ affects both the resubstitution and $K-fold$ losses in the KNN classifier, along with how quickly these losses change. The solid lines represent the actual losses, while the dashed lines show how the losses vary with each $K$ value. At $K=3$, we see that both loss values are minimized, which means this choice strikes a good balance between fitting the training data well and generalizing to new data. Choosing $K=1$ or $K=2$ slightly lowers the resubstitution loss, but it also raises the K-fold loss, indicating a risk of overfitting. The slopes are fairly stable around $K=3$, reinforcing this as the optimal choice to keep the model accurate without fitting the data too closely.

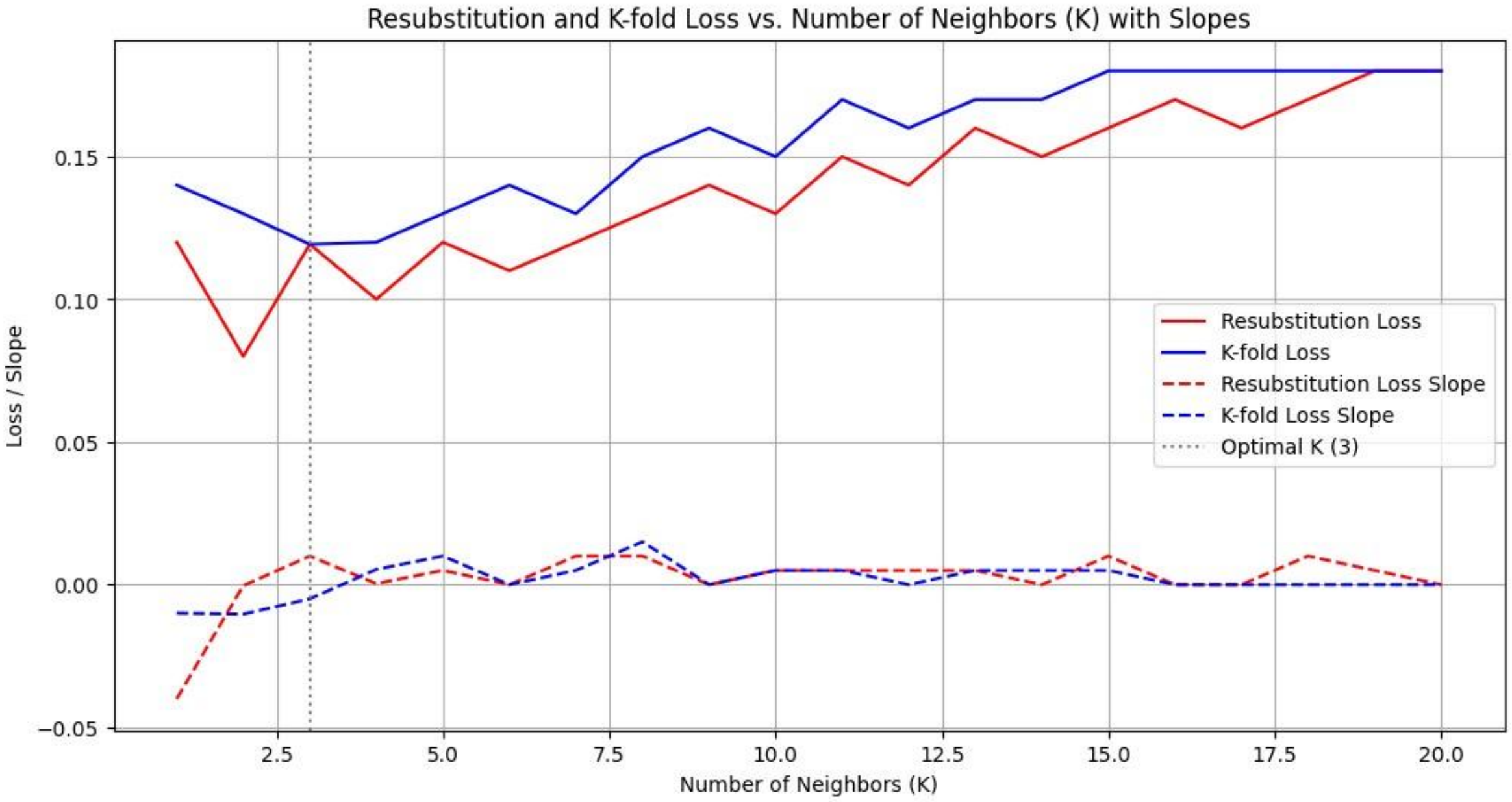

**Figure 9** demonstrates that $K = 3$ minimizes both the resubstitution and $K - fold$ losses, indicating an optimal balance between fitting the training data and generalizing well. Lower $K$ values (such as 1 or 2) reduce the resubstitution loss but increase the $K - fold$ loss, which risks overfitting. The slopes stabilize around $K = 3$, supporting it as the ideal choice for maintaining accuracy without overfitting.

### 3.4 Decision Tree

To effectively separate and distinguish classes, a decision tree or learning tree can be employed. A decision tree is a method used to estimate objectives with discrete values. It is one of the most well-known supervised learning algorithms and has been successfully applied across various domains due to its robustness against input data noise. Decision tree learning in this context utilizes the ID3 algorithm. Based on the decision tree classification, the strategy chosen is the one with the highest likelihood [22]. For distinguishing between faulty and fault-free systems using the decision tree method, output load current and battery charge state are used, similar to the PCA and KNN methods. **Table 4** presents the results of fault classification, including replacement fault rates and K-fold replication.

| | | | | | |
|---|---|---|---|---|---|
| **Resubstituting Loss** | **0.0137~ 1.3%** | | | | |
| **K_Fold Loss** | **0.0435~ 4.3%** | | | | |
| **Confusion Matrix** | **1938** | **0** | **31** | **5** | **27** |
| | **0** | **2001** | **0** | **0** | **0** |
| | **24** | **0** | **1968** | **3** | **6** |
| | **0** | **0** | **4** | **1997** | **0** |
| | **30** | **1** | **6** | **0** | **1964** |

**Table 4** decision tree calcification result

**Table 4** indicates that, similar to the KNN method, the Decision Tree (DT) method does not achieve complete fault classification. According to this table, the discrete accuracy of the classes is as follows: the fault-free system class has an accuracy of 97%, the battery ground fault class has 99.99%, the open circuit fault in the IGBT at the MPPT converter has 98%, and the open circuit fault and IGBT short circuit in the regulator converter range from 99.6% to 98.35%. These results demonstrate that the DT method shows particularly high accuracy in classifying battery ground faults and short circuit faults in the IGBT regulator converter.

### 3.5 PCA Method

The Principal Component Analysis (PCA) method provides a linear mapping between inputs and outputs, with the discrete system being based on their eigenvalues and eigenvectors. **Figure 10** presents the input plane of output load current and battery charge state across system fault classes and potential fault types within the system, shown in two dimensions before applying the PCA method.

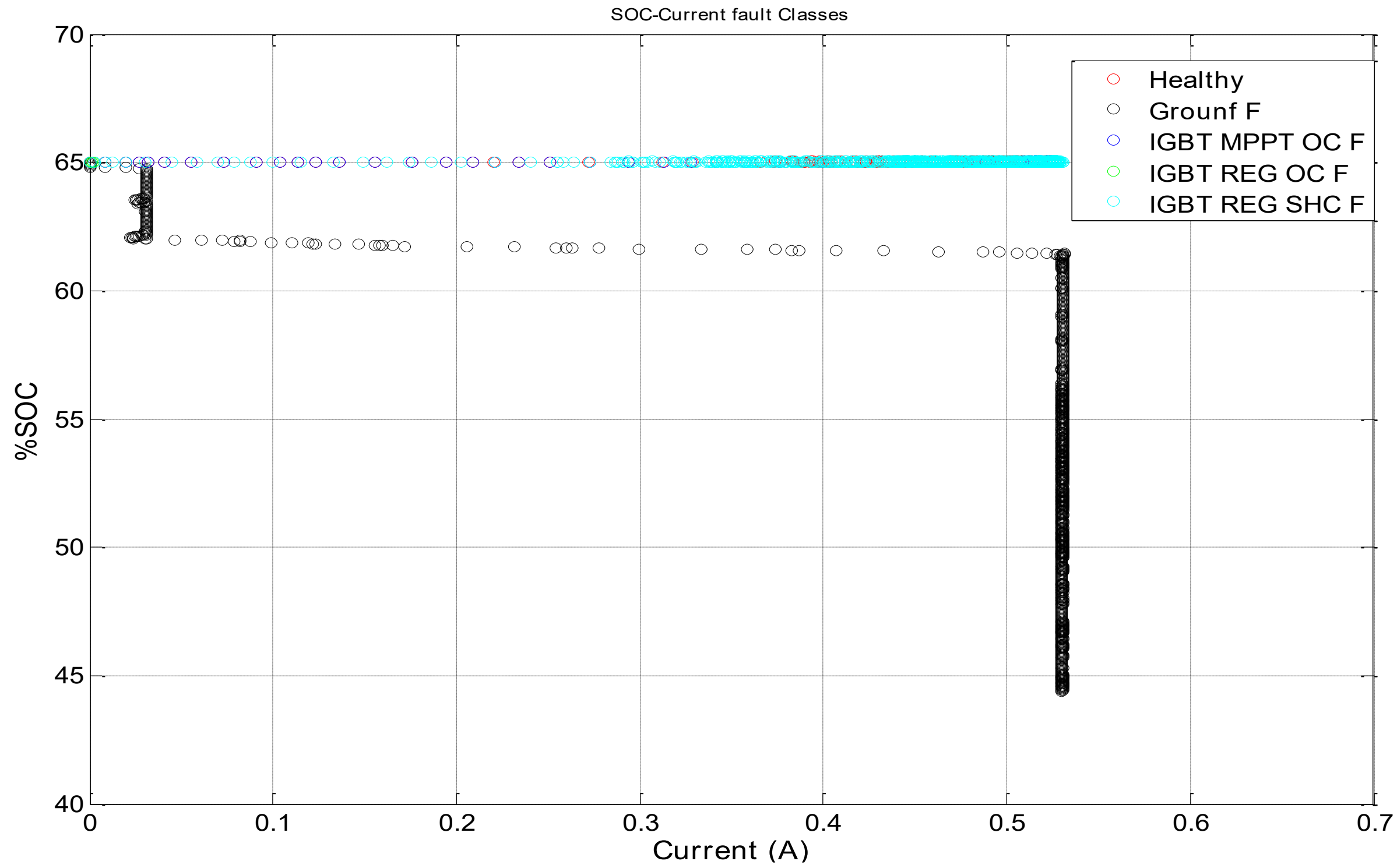

**Figure 10** Fault types on the plane of battery charge state and load current.

**Figure 10** illustrates that the states of fault-free conditions and potential fault classes within the electrical power system cannot be distinctly separated on the two-dimensional plane of battery charge state and output load current. For using the PCA method, first of all, inputs classifiers are Ordered pair ($I_{L_i}$,$SOC_i$) and matrix X at **Equation 9** is considered.

$$y = Q^T.X \tag{9}$$

In **Equation 10**, y represents the mapped output vector, and Q is the eigenvector matrix corresponding to the covariance matrix C in **Equation 10**. In this equation, the mean of the data (bias) is assumed to be zero, as defined below:

$$C = Cov(X) = E\{(X - X^T)(X - X^T)\} \tag{10}$$

To calculate the PCA method, the matrix Q, which contains the eigenvectors corresponding to the largest eigenvalues of the covariance matrix C, is determined. The eigenvectors are ordered in ascending order by their associated eigenvalues in the columns of matrix Q. The eigenvectors with the largest eigenvalues are then selected, as they produce the greatest variance dispersion in the y output. The results of fault separation and classification using the PCA method, including the fault-free state and specific fault classes, are presented in **Figure 11**.

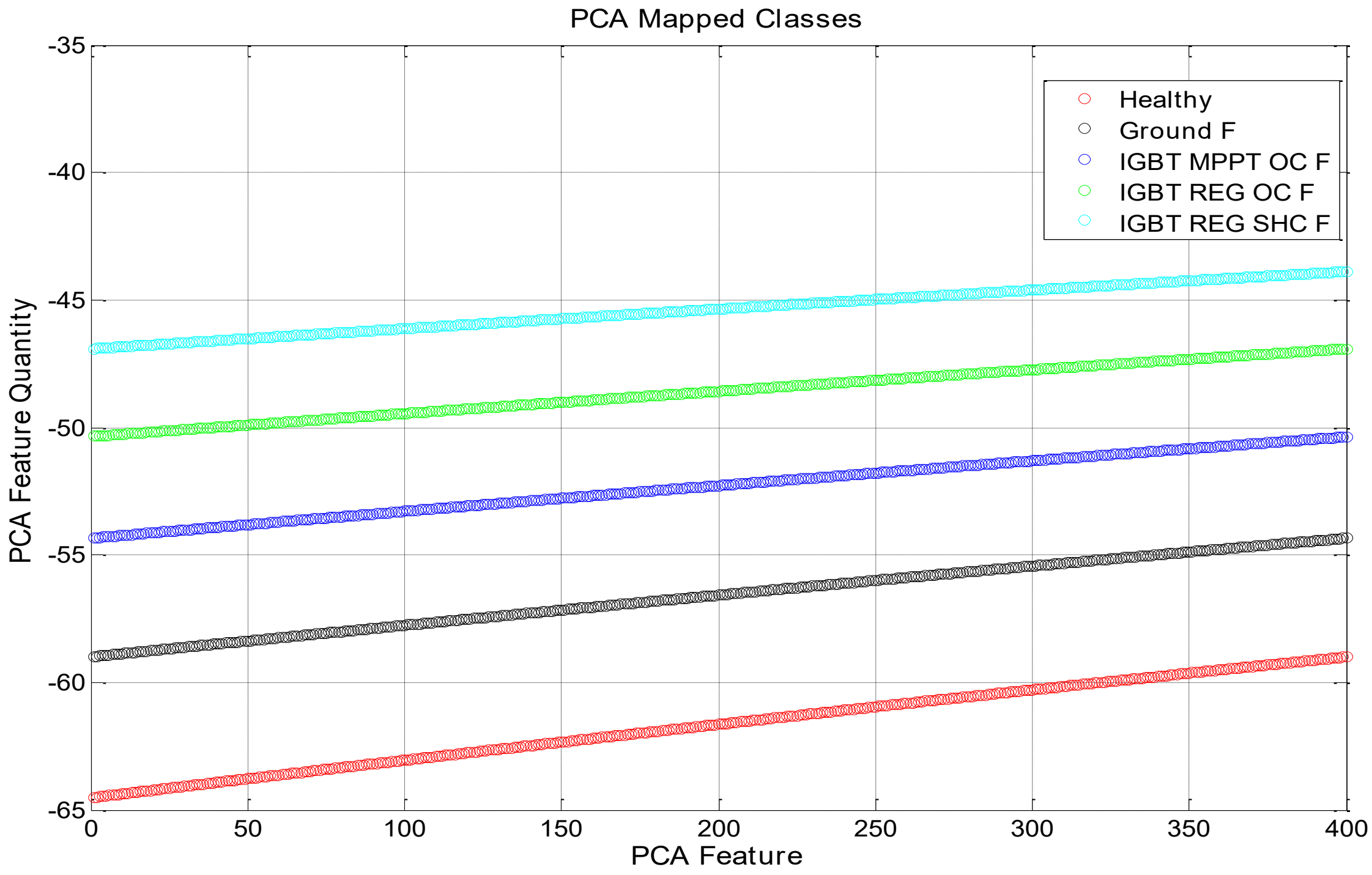

**Figure 11** Separation of possible faults in the electrical power system using the PCA method.

**Figure 11** demonstrates that both the fault-free state and potential fault types are reduced to one dimension with appropriate spacing and are effectively separated using the PCA method. The results and conclusions of the methods used for separating faults in the electrical power system, excluding the photovoltaic subsystems, are summarized in **Table 5**.

| Method | Accuracy(%) | System |
|---|---|---|
| **Neural network (MLP)** | 99.5 | Electrical power |
| | 99.2 | photovoltaic |
| **PCA** | 99.5 | Electrical power |
| **KNN** | 93 | Electrical power |
| **DT** | 98.6 | Electrical power |

**Table 5** compares classification methods for electrical power system faults.

**Table 5** shows that the accuracy of the neural network MLP and PCA methods, which are two algorithms used for separating and classifying possible faults in the satellite's electrical power system, is higher than that of other methods. The Decision Tree (DT) method also demonstrates high accuracy, while the KNN method, based on its accuracy, is not suitable for this purpose. For a detailed analysis of the classification quality of possible faults in the electrical power system within each class, refer to **Table 6**.

| | Without fault | Battery ground | Open circuit IGBT MPPT | Open circuit IGBT regulator | Shor circuit IGBT regulator |
|---|---|---|---|---|---|
| **MLP1** | 45.2 | 0.6 | 23.3 | 2.5 | 30 |
| **MLP2** | 0 | 0 | 0 | 0 | 0 |
| **KNN** | 14 | 0 | 13 | 0.1 | 0.7 |
| **TD** | 2.7 | 0 | 2 | 0.4 | 1.6 |
| **PCA** | 0 | 0 | 0 | 0 | 0 |

**Table 6** compares the accuracy of fault classification in the electrical power system.

**Table 6** indicates that the battery ground fault has a higher likelihood of occurrence compared to other classes, as identified by the methods presented. Additionally, separating the fault-free system class proved to be more challenging and less accurate.

## 4. Conclusion

This paper proposed a simulation of the electrical power system in a satellite. The simulation considered effective parameters influencing the power system's performance, with inputs as parameters and the electrical load current as the output, accounting for mass and volume limitations. By considering these parameters and applying reliability analysis using neural networks on system equipment, the electrical system model, both in fault-free and possible fault states, was determined with suitable accuracy. Subsequently, classification methods including neural network MLP, PCA, KNN, and DT were employed to separate fault classes. The results demonstrated that the neural network MLP and PCA methods were more effective in separating and classifying fault-free and possible fault classes, outperforming the KNN and DT methods.

Another key focus will be to make the model more adaptable, so it can detect new types of faults as satellite technology advances. By incorporating additional machine learning techniques such as Reinforcement Learning or exploring hybrid models, improve its ability to handle new and unexpected fault scenarios. Additionally, this approach could be extended to other satellite subsystems, not just the electrical power system, to create a more comprehensive diagnostic tool. Partnering with industry to collect diverse data and test the model in real satellite operations will also be essential. In the long run, these steps will help develop a fault detection system that's versatile and reliable enough to keep up with the changing demands of satellite mission.